\documentclass[a4paper,showkeys,floatfix,aps,pre,preprint,groupedaddress]{revtex4-1}
\usepackage{graphicx} % Required for inserting images
\usepackage{amsmath, amssymb}  % For equations
\usepackage{hyperref}    % For hyperlinks
\usepackage{geometry}    % For adjusting page layout
\usepackage{natbib}      % For bibliography
\usepackage{xcolor}
\begin{document}

\title{Does Excellence Correspond to Universal Inequality Level?}
\author{Soumyajyoti Biswas}
\affiliation{Department of Physics \& Department of Computer Science and Engineering, SRM University - AP, Andhra Pradesh 522240, India}
\author{Bikas K. Chakrabarti} 
\affiliation{Saha Institute of Nuclear Physics, Kolkata 700064, India}
\affiliation{Economic Research Unit, Indian Statistical Institute, Kolkata 700108, India.}
\author{Asim Ghosh}
\affiliation{Department of Physics, Raghunathpur College, Raghunathpur, Purulia 723133, India}
\author{Sourav Ghosh}
\affiliation{Department of Physics, SRM University - AP, Andhra Pradesh 522240, India}
\author{M\'at\'e J\'ozsa}
\affiliation{{Babe\c{s}-Bolyai University, Deptartment of Physics, Cluj-Napoca, Romania}}
\author{Zolt\'an N\'eda}
\affiliation{{Babe\c{s}-Bolyai University, Department of Physics, Cluj-Napoca, Romania}}
%\date{February 2025}

\begin{abstract}
    We study the inequality of citations received for different publications of various researchers and Nobel laureates in Physics, Chemistry, Medicine and Economics using Google Scholar data from 2012 to 2024. Citation distributions are found to be highly unequal, with even greater disparity among Nobel laureates. Measures of inequality, such as the Gini and Kolkata indices, emerge as useful indicators for distinguishing Nobel laureates from others. Such high inequality corresponds to growing critical fluctuations, suggesting that excellence aligns with an imminent (self-organized dynamical) critical point. Additionally, Nobel laureates exhibit systematically lower values of the Tsallis--Pareto parameter \( b \) and Shannon entropy, indicating more structured citation distributions. We also analyze the inequality in Olympic medal tallies across countries and find similar levels of disparity. Our results suggest that inequality measures can serve as proxies for competitiveness and excellence. 
\end{abstract}

\keywords{Citation data; Inequality indices; Lorenz curve; Gini index; Kolkata index}

\maketitle

\section{Introduction}

The inequality of acquired resources among competing entities is a well-established empirical observation. It is particularly well studied in the context of income and wealth inequalities~\cite{stauffer}. More than a century ago, Pareto formulated, based on empirical data, his  80 -- 20 %MDPI: Please check if this should be endash.
 law, which states that 20\% of people in a society posses 80\% of the total wealth~\cite{pareto}. It has since been modified and used in a wide variety of contexts from managements  (see, e.g.,~\cite{manage}) and infectious disease spreading~\cite{gu1,gu2} to physical systems undergoing (or are near) a phase transition (see, e.g.,~\cite{manna}). More recently, this principle has also been applied to scenarios involving asset accumulation among many competing entities, where such competitions are seemingly unrestricted. Unlike cases such as wealth distribution or disease spread, where inequality often signals underlying issues, in these scenarios, inequality emerges naturally and can even be desirable. Such examples include the inequality of the fractions of votes received by candidates in an election, the inequality of the incomes of producers from different movies~\cite{ijmpc}, and also the inequality of scholarly citations of the publications of a researcher~\cite{zoltan,nielsen}.
{In particular, the distribution of citations among an individual’s publications has been found to follow a heavy-tailed form. Prior work suggests that the Tsallis--Pareto (also known as Lomax II) distribution provides a good fit to the empirical citation data of many authors~\cite{zoltan}. Earlier studies also noted the difference in citation patterns between successful researchers and others~\cite{dong}. However, the reasons for citation inequality is sometimes also contributed by structural biases (see, e.g.,~\cite{teich,nett}) and field-dependent citation patterns (see, e.g.,~\cite{crespo}). At the level of journals, such inequality will affect the impact factors~\cite{kiess}.}
Given these contexts, there is almost no need for an external intervention in mitigating the emergent inequalities in these cases, hence the competitions can be allowed to evolve in an unrestricted manner. 

On the other hand, it has also been suggested---within a limited scope---that scientific excellence, when defined through prizes and awards of high repute, are often accompanied by a high level of inequality in scholarly citations of different publications of the recipient of such prizes or awards~\cite{entropy,Qfactor}. It has been a long-standing quest to obtain a reliable metric to asses scientific excellence. In this work, we study the emergent inequality in scholarly citations and
seek at least a robust correlation between scientific excellence (measured through wider recognitions) and an inequality of scholarly citations of the different publications of a researcher. Using Google Scholar citation data for 126,067 researchers (each having at least 100 publications), we show that measures of inequality of citations could be a useful indicator of excellence.  

We also look at another system involving unrestricted competition, which is the distribution of medals in Olympic games. We show that a very similar nature of inequality exists even for this case.   
%%%%%%%%%%%%%%%%%%%%%%%%%%%%%%%%%%%%%%%%%%%%%%%%%%%%%%%%%

%%%%%%%%%%%%%%%%%%%%%%%%%%%%%%%%%%%%%%%%%%
\section{Methods}
\begin{figure}
\includegraphics[width=0.99\linewidth]{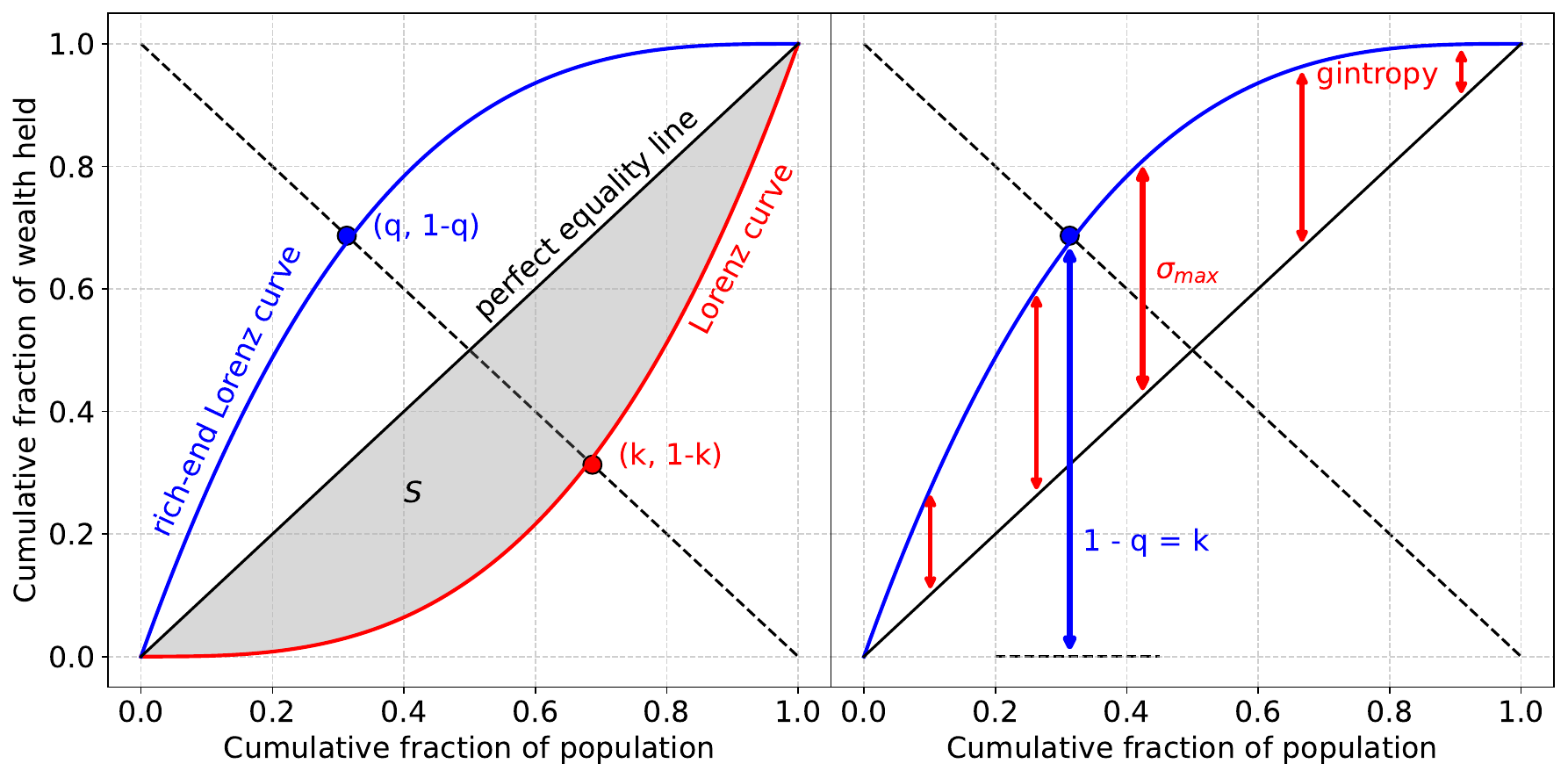}
	\caption{{Different %MDPI: 1. Please change the hyphen (-) into a minus sign ($-$, "U+2212"). e.g., "-1" should be "$-$1". 2. Please add the explanation of dot line. 3. Please add the explanation of red arrow in right of Figure. 
 types of Lorenz curves and their relation to the studied measures: Gini index ($g$), Kolkata index ($k$), Pareto index ($q$), and maximum gintropy ($\sigma_{max}$).
			(\textbf{Left}}) The Lorenz curve (in red) denotes the cumulative fraction of wealth or ``assets'' held by a cumulative fraction of the population when arranged in ascending order of their wealth. The rich-end Lorenz curve (shown in blue) is the same when the population is arranged in descending order of their wealth. If all agents have exactly the same wealth, then the Lorenz curve is a 45-degree straight line, called the perfect equality line. The area ($S$) between this line and the Lorenz curve (shaded region) is then one measure of inequality. Normalizing this area by the total area under the perfect equality line yields the Gini index ($g=2S$). The off-diagonal intersects the Lorenz curve at $(k,1-k)$ and the rich-end Lorenz curve at $(q,1-q)$, where $1-k$ fraction of the population holds $k$ fraction of the total wealth, defining $k$ as the Kolkata index. In terms of the Pareto index, a fraction $q$ of the population possesses $1-q$ fraction of the wealth, implying $q=1-k$. {(\textbf{Right}) Illustration of the ``gintropy'' measure, defined as the density function given by the difference between the rich-end Lorenz curve and the perfect equality line (green arrows %MDPI: No green arrows in the figure, please check.
). The maximum value of this function is denoted as $\sigma_{max}$, which, as shown in the figure, exhibits a strong correlation with $1 - q = k$, the Kolkata index.}}
	\label{lorenz_curve}
\end{figure}

As mentioned above, we quantify the inequality of ``assets'', which we use as a generic term, in order to correlate with individual successes. Such quantification mechanisms already exist in the context of the quantification of wealth inequality. Particularly, if a total asset $A$ is distributed among a group of $N$ individuals, where the $i$-th individual has an $a_i$ amount of assets ($\sum\limits_{i=1}^Na_i=A$), then one can first arrange the individuals in the ascending order of the assets possessed by them. Then, it is possible to define a Lorenz curve $L(p)$~\cite{lorenz} that indicates the fraction of the total assets possessed by the poorest (in terms of the particular type of asset concerned) $p$ fraction of the individuals. As is evident, the two limiting points of the Lorenz curve would be $L(0)=0$ and $L(1)=1$,{ with a monotonically increasing function in between.} %EE: Please check intended meaning is retained.­
 Clearly, if each individual had exactly equal assets ($A/N$), then the Lorenz curve would be a 45-degree straight line. This line represents perfect equality. On the other hand, any inequality in the asset distribution would necessarily make the curve concave, and the area opening up between the equality line and the actual Lorenz curve would then be a measure of inequality in the asset distribution (see Figure~\ref{lorenz_curve}).  Indeed, it is this area normalized by the area under the equality line that is defined as the Gini index~\cite{gini}, which is widely used and interpreted in Economics and various other \mbox{fields~\cite{gu1,gu2,Bir2020}} as a common measure of inequality. The mathematical definition could be written as \linebreak $g=1-2\int\limits_0^1L(p)dp$. A discrete version for the Gini index can also be \mbox{similarly defined. }

Another measure of inequality that we use here is the Kolkata index ($k$)~\cite{kolkata} (see also,~\cite{muthu}), which is a generalization of the Pareto's law. It is the fixed point of the Complementary
Lorenz function $\tilde{L} (p)  (\equiv 1 -  L(p)): \tilde{L}(k) = k$.  As such, it says that $1 - k$ fraction
of the richest individuals (persons/papers/countries, \ldots) possess $k$ fraction of the
total assets (wealth/citations/medals, \ldots  respectively; see Figure~\ref{lorenz_curve}). Clearly, in the limit $k=0.8$, one obtains the 80 -- 20 law. This index complements the Pareto index $q$, which describes how the wealth is distributed at different levels: for any fraction $q$, the richest $q$ portion of the population owns $1-q$ of the total wealth. Extending this iteratively, the richest $q^n\times 100\%$ holds $(1-q)^n\times 100\%$ of the total wealth, with $n$ being a non-negative real~\cite{pindex}.   Unlike the Gini index, the Kolkata index  has a more intuitive interpretation related to the reflection of the underlying inequality at least at its extreme end{, i.e.,} %EE: Please check intended meaning is retained.­
the richest individuals (see, e.g.,~\cite{kpa1,kpa2} for its application in kinetic wealth exchange model). 
{The Kolkata index ($k$) exhibits a strong correlation with the maximum value of the “gintropy” measure ($\sigma_{\max}$) (see Figure~\ref{lorenz_curve}). Gintropy itself is a density-like quantity, defined as the difference between the rich-end Lorenz curve and the perfect equality line (diagonal). Introduced in%MDPI: We reorder the reference number, please check and confirm.
~\cite{Bir2020}, gintropy establishes a connection between the Gini index and entropy, as implied by its name. Its entropy-like density characteristics were demonstrated, suggesting that the Kolkata index ($k$), given its high correlation with $\sigma_{\max}$, can be interpreted as an inequality measure with entropic properties.
}

%\vspace{-3pt}

{To interpret the structure of asset distributions using information-theoretic measures, we considered Shannon entropy for continuous probability density functions (PDFs). For a PDF \( \rho(x) \), it is defined as \( S = -\int \rho(x) \log \rho(x) \, dx \), capturing the uncertainty inherent in the distribution.}

{We analyzed individual citation distributions by fitting the Tsallis--Pareto (Lomax II) distribution, which has previously been shown to capture such data accurately~\cite{zoltan}. The rescaled one-parameter probability density function is defined as
	\[
	\rho(x) = \frac{b}{b - 1} \left(1 + \frac{x}{b - 1} \right)^{-1 - b},
	\]
	where \( x \) is the citation count and \( b \) is the shape parameter. The corresponding cumulative distribution function,
	\[
	F(x) = 1 - \left(1 + \frac{x}{b - 1} \right)^{-b},
	\]
	was used to ensure a smooth empirical fit.
	For each researcher, the optimal \( b \) was obtained by scanning over a discrete set of values in the interval \( b \in (1, 8) \), and minimizing the average of squared relative logarithmic differences between the empirical and theoretical cumulative distributions:
	$ \frac{1}{N} \sum_{i=1}^{N} \left( \frac{\log F_i^{\text{exp}} - \log F_i^{\text{theo}}}{\log F_i^{\text{exp}}} \right),$
	where \( N \) is the number of data points. The fit was performed in logarithmic scale to accurately capture the tail behavior. This method was applied to all researchers in the dataset, including Nobel laureates, allowing for a comparison of the fitted \( b \)-parameters across groups.
}

Along the line of exploring the effect of the richest individuals in determining the inequality, we also looked at a quantity $Q$-factor~\cite{Qfactor}, which is the ratio of the maximum value of asset possessed by the (richest) individual and the average asset value: \linebreak $Q=max\{a_i\}/(A/N)$. In terms of citations, for example, this will be the ratio of the citation number of the highest cited paper of an individual and the average citation of all papers of that individual. Similar extensions can be made for Olympic medals as well. 

%%%%%%%%%%%%%%%%%%%%%%%%%%%%%%%%%%%%%%%%%%%%%%%%%%%%%%%%%

%%%%%%%%%%%%%%%%%%%%%%%%%%%%%%%%%%%%%%%%%%%%%%%%%%%%%%%%%%

Along with these, we also continued to use the more commonly used (for the case of citations), the Hirsch index ($h$)~\cite{hindex} that states $h$ publications of the concerned individual have at least $h$ citations each {(see~\cite{h_rev1,h_rev2} for a review and other variants of the $h$ index)}.   

Now, the inequality indices are defined in a generic way such that it is possible to use these in a wider context. Particularly, we will focus on the inequality of citations of the publications of a given researcher and inequality of the medals received by different countries in the Olympic games (both summer and winter versions). As mentioned in the Introduction, both of these scenarios are examples of unrestricted competitions. We used Google Scholar data for 126,067 scientists (with more than 100 papers each) and \mbox{80 Nobel} laureates in different fields during the period 2012--2024. The datasets are available in~\cite{data1}.
We also used the data for the Olympic medals received by different countries, both in the summer and winter versions of the games within the period 1896--2024 and 1924--2022, respectively; the data are available in~\cite{data2}. 
%%%%%%%%%%%%%%%%%%%%%%%%%%%%%%%%%%%%%%%%%%%%%%%%%%%%%%%%%

%\end{document}

%%%%%%%%%%%%%%%%%%%%%%%%%%%%%%%%%%%%%%%%%%

%%%%%%%%%%%%%%%%%%%%%%%%%%%%%%%%%%%%%%%%%%
\section{Results}

\subsection{Inequality in Citations}
We started with investigating the inequality in the citation data of the publications by individual researchers. In particular, we considered the citations received {from} %EE: Please check intended meaning is retained.­
the various papers of a researcher (including papers with no citations) who have at least 100 publications to ensure sufficient statistics, and measured the indices $g$, $k$, $h$ and $Q$ as described earlier. We carried out the same exercise for Nobel laureates in Physics, Chemistry, Medicine and Economics during the years  2012--2024, for the individuals with a public Google Scholar profile. 

There are some prior observations that we need to first discuss in this context. It has been noted elsewhere that the citation inequality of different papers of a researcher, quantified through the Gini index, is surprisingly high~\cite{zoltan}. However, if the successful researchers were considered (in terms of Nobel laureates, Fields Medalists, Boltzmann awardees, etc.), the inequality is, in general, even higher~\cite{ijmpc}, indicated by both the Gini and Kolkata indices. Particularly, even though the ranges of these two indices are not the same [(0, 1) for $g$ and (0.5, 1) for $k$], they tend to become equal (close to 0.87) for successful researchers.
Firstly, this would mean that the citation inequality is more prominent among successful researchers. This observation could then be translated as indicators of excellence. Secondly, it is worth mentioning that the tendency of $g$ and $k$ to fluctuate around 0.87 has been found in studying the inequality of responses in many physical systems at or near a critical point~\cite{manna}. It has been shown analytically and numerically that the crossing point of $g$ and $k$ signals an imminent system spanning response and that the value at the crossing point is rather weakly dependent on the underlying distribution function (assumed to be power law near the critical point). This in turn would then suggest that successful researchers could be near a self-organized critical state~\cite{manna,soc-citation}. {This is, of course, based on observational similarity with the behavior of known SOC systems.}

%%%%%%%%%%%%%%%%%%%%%%%%%%%%%%%%%%%%%%%%%%%%%%%%%%%%%%%%%

%%%%%%%%%%%%%%%%%%%%%%%%%%%%%%%%%%%%%%%%%%%%%%%%%%%%%%%%%%

\begin{figure}
	
	\includegraphics[width=0.75\linewidth]{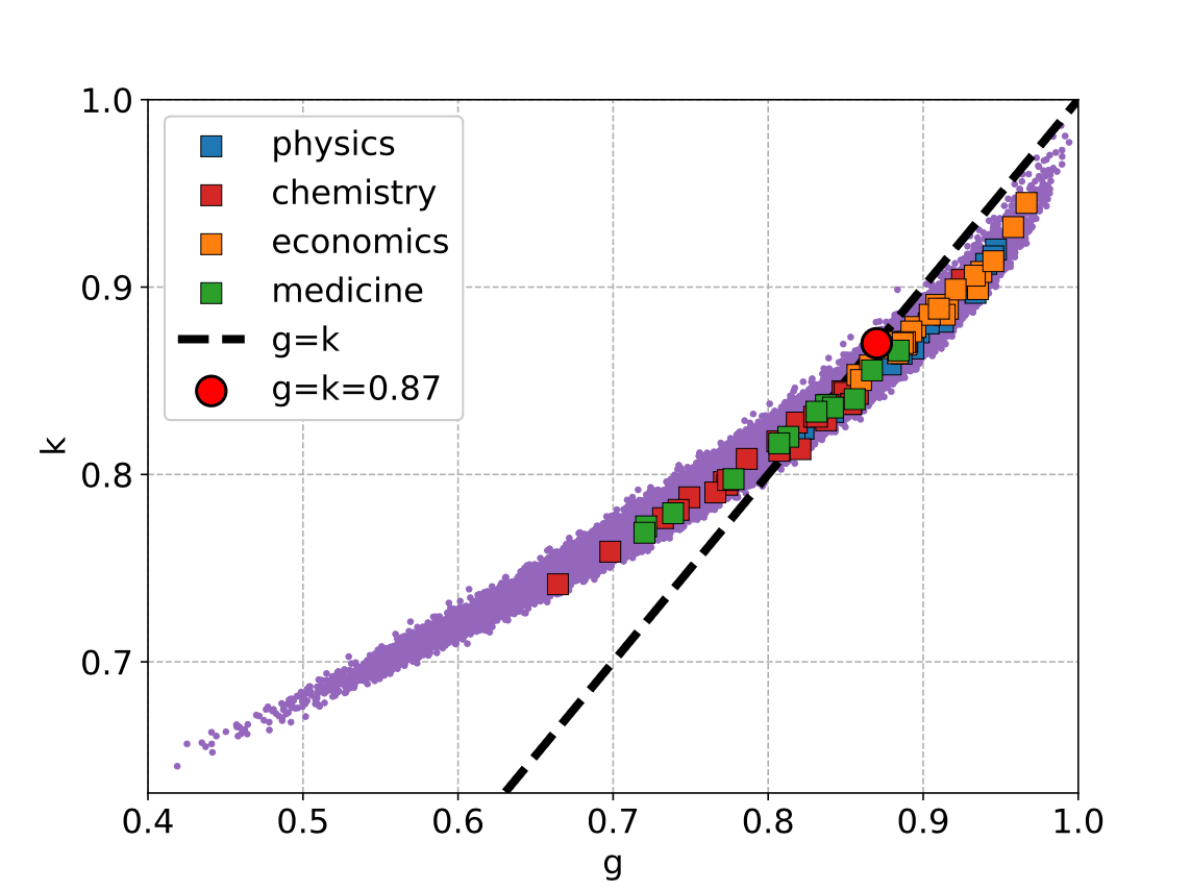}
	\caption{Illustration %MDPI: 1. We have moved the position of the Figures 2--8. Please confirm. Please revise and ensure the first citation of each Figure appears in numerical order. 2. Please add the explantion of purple.
 of %MDPI: 
 citation inequality among an individual researcher's papers, quantified using the Gini index ($g$)~\cite{gini} and the Kolkata index ($k$)~\cite{kolkata}. The background data (purple dots) represent 126,067 scientists with more than 100 published papers, based on Google Scholar records. Nobel laureates from 2012 to 2024 across different disciplines are highlighted using distinct colors. The dashed line corresponds to $g=k$, around which the data points for Nobel laureates tend to cluster. A seemingly universal critical point at $g=k=0.87$ is marked by a red dot. Note that $k = 0.8$  for any author would suggest  that
80\% of citations  come from 20\% of the papers by that
author (Pareto law).}
	\label{k_vs_g}
\end{figure}

In view of the above, we first plot $g$ vs. $k$ in Figure~\ref{k_vs_g} for all  126,067 scientists and the \mbox{80 Nobel} laureates in the abovementioned fields during the period 2012--2024. We also show the $g=k$ line for reference. As can be seen, the data for the Nobel laureates are clustered around $g\approx k$. This clustering, observed against the broader distribution of researchers, suggests that these inequality measures may serve as potential indicators for distinguishing highly successful researchers from the general population.

\begin{figure}
 \includegraphics[width=\linewidth]{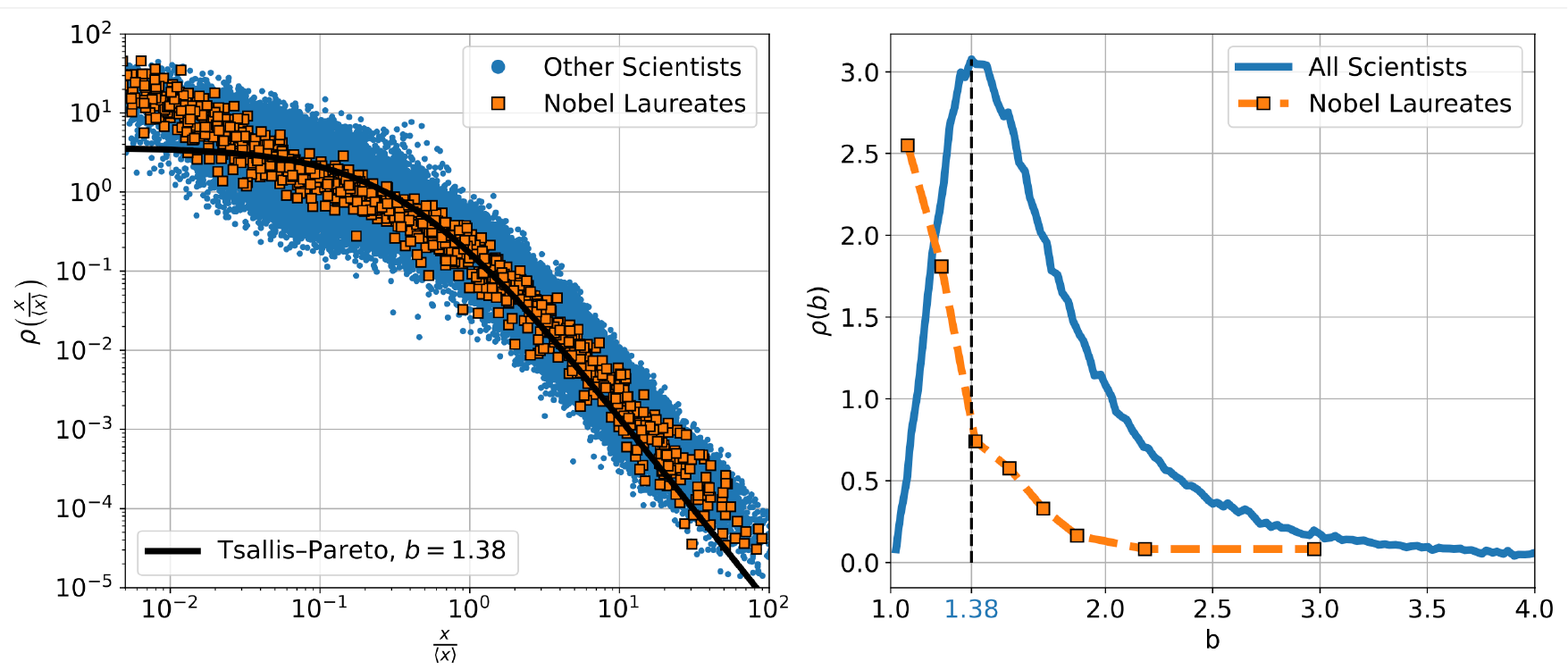}
	\caption{Analysis of Tsallis--Pareto fits to citation count data for Nobel laureates and the general scientific population.	(\textbf{Left}) Citation distributions for all Nobel laureates (orange squares) and a random sample of 20,000 scientists from the general population (blue dots). The distributions overlap significantly and follow the same Tsallis--Pareto form, with the most probable fit corresponding to \( b = 1.38 \) (see right panel). The Nobel laureates form a slightly narrower band of distributions. \mbox{(\textbf{Right}) Distribution} of fitted Tsallis--Pareto parameters \( b \) for the general population (blue) and Nobel laureates (orange). Nobel laureates tend to exhibit lower \( b \) values, clustering closer to \( b \to 1 \). The peak at \( b = 1.38 \) is indicated in blue.}
	\label{b_fit}
\end{figure}

\begin{figure}
	\includegraphics[width=0.99\linewidth]{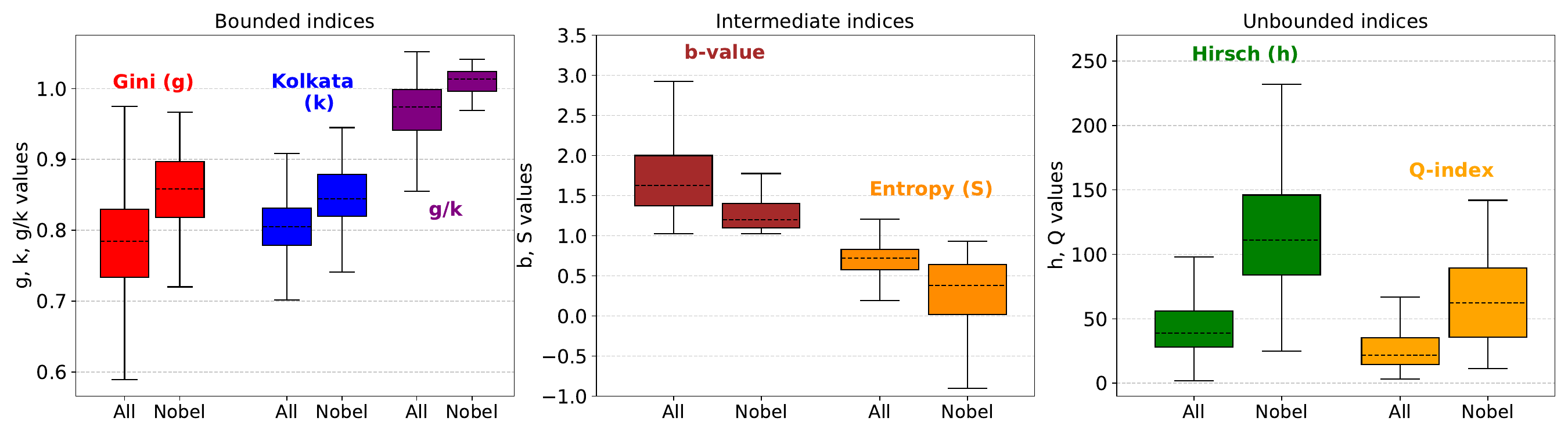}
	\caption{Box plot comparison of citation-based metrics for all researchers and Nobel laureates. The left panel shows bounded indices: Gini ($g$, red), Kolkata ($k$, blue), and their ratio $g/k$ (purple), constrained to $g \in (0,1)$, $k \in (0.5,1)$, and $g/k \in (0,2)$. The middle panel presents the Tsallis--Pareto shape parameter $b$ (green%MDPI: No Green in middle panel, please check and revise.
), defined on $(1,\infty)$, and the Shannon differential entropy $S$ (orange), defined on $(-\infty,\infty)$. While formally unbounded, these indices span intermediate ranges in practice, narrower than $h$ and $Q$. The right panel shows the unbounded indices: Hirsch index ($h$, green) and Q-factor (orange), both on $(0,\infty)$. Nobel laureates exhibit systematically lower values of $b$ and $S$, and higher values in all other indices, compared to the general population. The separation is evident across all panels, supporting the relevance of these metrics in distinguishing scientific excellence.
	}
	\label{boxplot}
\end{figure}

We examined the citation distributions of Nobel laureates using the Tsallis--Pareto form introduced earlier. Their empirical distributions are well described by this model, similarly to the general population (see left panel of Figure %MDPI: Wrong Figure ciation number: Figure 5”, “prev number is Figure 2”. Figure 2’s ciation should before Figure 3’s ciation. Please revise and ensure the first citation of each Figure appears in numerical order. Same as below.
\ref{b_fit}). However, a notable distinction lies in the fitted values of the shape parameter \( b \), which, for Nobel laureates, are systematically lower and tend to cluster near the lower bound \( b \to 1 \) (see also the middle panel of Figure%MDPI: %MDPI: Wrong Figure ciation number: Figure 6”, “After number is Figure 3”. Please revise and ensure the first citation of each Figure appears in numerical order. Same as below.
~\ref{boxplot}). As shown in the right panel of Figure~\ref{b_fit}, the distribution of \( b \) values for Nobel laureates is clearly shifted compared to that of other researchers. This suggests that the parameter \( b \) may also serve as a useful indicator of scientific excellence, with Nobel laureates appearing close to the lower limit of the typical citation distribution.

%\vspace{-3pt}

\begin{figure}

\includegraphics[width=0.75\linewidth]{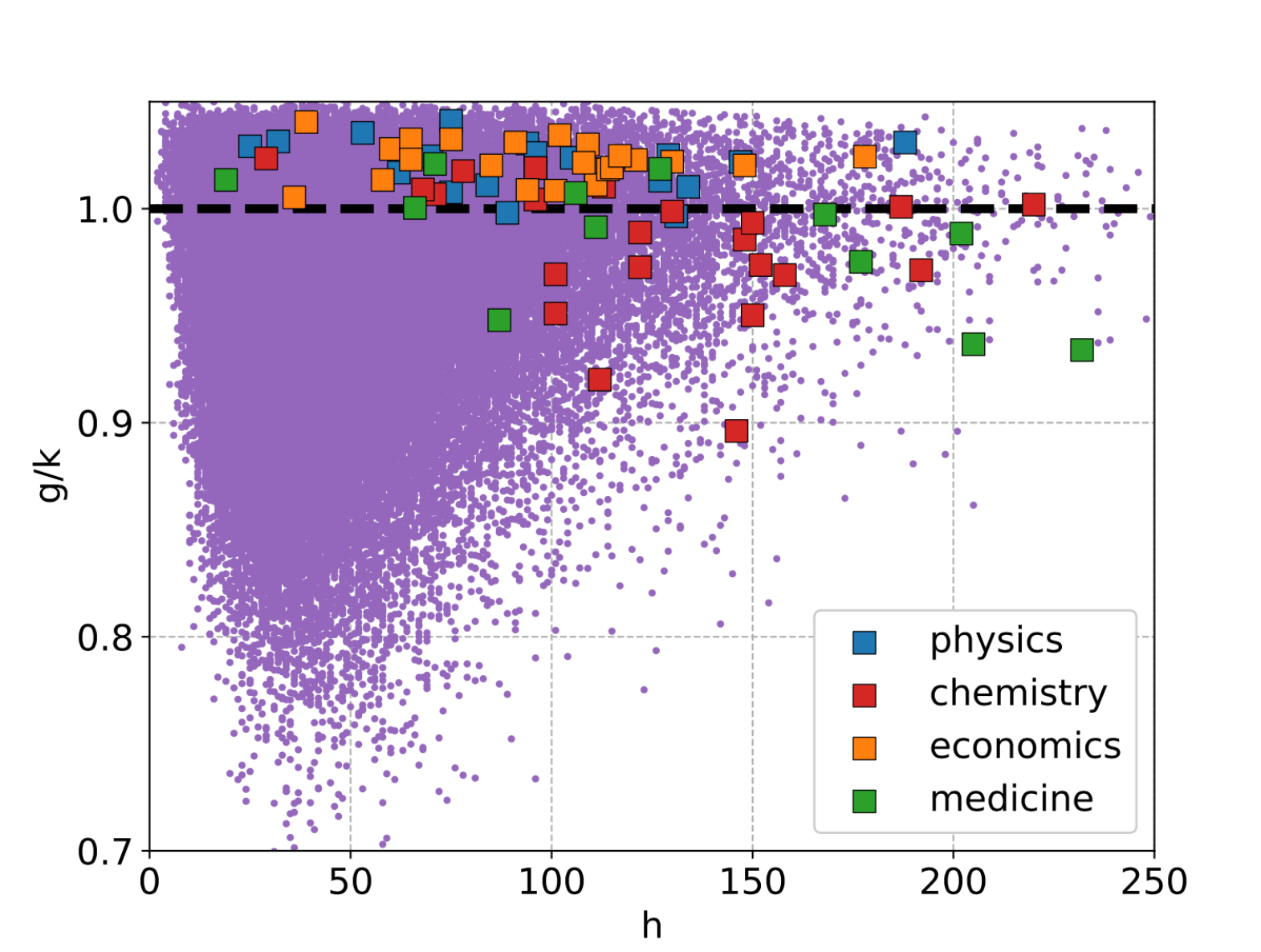}
\caption{Illustration %MDPI: We have moved the position of the Figures 2--8. Please confirm. Please revise and ensure the first citation of each Figure appears in numerical order. Same as Figures 2--8. 2. Please add the explantion of purple.
 of the ratio between the Gini index ($g$)~\cite{gini} and the Kolkata index ($k$)~\cite{kolkata}, representing citation inequality among individual researchers' papers, plotted as a function of the Hirsch index ($h$)~\cite{hindex} for 126,067 researchers (purple dots). Nobel laureates from different disciplines are highlighted using distinct colors. All data are sourced from Google Scholar. The dashed line represents $g/k = 1$, around which the Nobel laureate data tend to cluster (mostly within a range $1.00\pm0.03$; see also Figure~\ref{boxplot}). Notably, data points from Physics and Economics tend to group together, distinct from those of Chemistry and Medicine. The former cluster generally exhibits $g/k>1$ with slightly lower $h$ values, whereas the latter shows $g/k < 1$ with comparatively higher $h$ values.}
\label{h_vs_gk}
\end{figure}

It is worth discussing at this point the effectiveness of the well-known Hirsch index in making such distinctions. Particularly, in Figure~\ref{h_vs_gk}, we plot $h$ on the x-axis and $g/k$ values on the y-axis. The Nobel laureates, as before, are indicated separately. The first pint to note is that the Nobel laureates have $h$-index values that are widely spread, but in terms of $g/k$, the ranges are narrow (see also left panel of Figure~\ref{boxplot}) and close to unity. There is, of course, the point that the $h$-index does not have an upper bound as such; indeed, it is often claimed to be related directly to the total number of citations ($N_c$) as $h\sim \sqrt{N_c}$~\cite{yong}. It is a monotonically increasing quantity with time (unlike $g$ and $k$) and often tends towards a high value for Nobel laureates (possibly due to a significant increase in $N_c$ after winning the prize). The second point here is that there is a clear abundance in the number of researchers near $g\approx k$ for whom the $h$ index is higher. This might qualitatively indicate that, after all, such a correlation would imply placing high values of $h$ in a similar footing as $g\approx k$. 
Thirdly, an interesting observation is the clustering of Physics and Economics researchers together, distinct from the Chemistry and Medicine groups. The former predominantly exhibit $g/k > 1$ with lower $h$-index values, whereas the latter tend to have $g/k < 1$ and higher $h$-index values. 

%\vspace{-6pt}
\begin{figure}
	
	\includegraphics[width=0.99\linewidth]{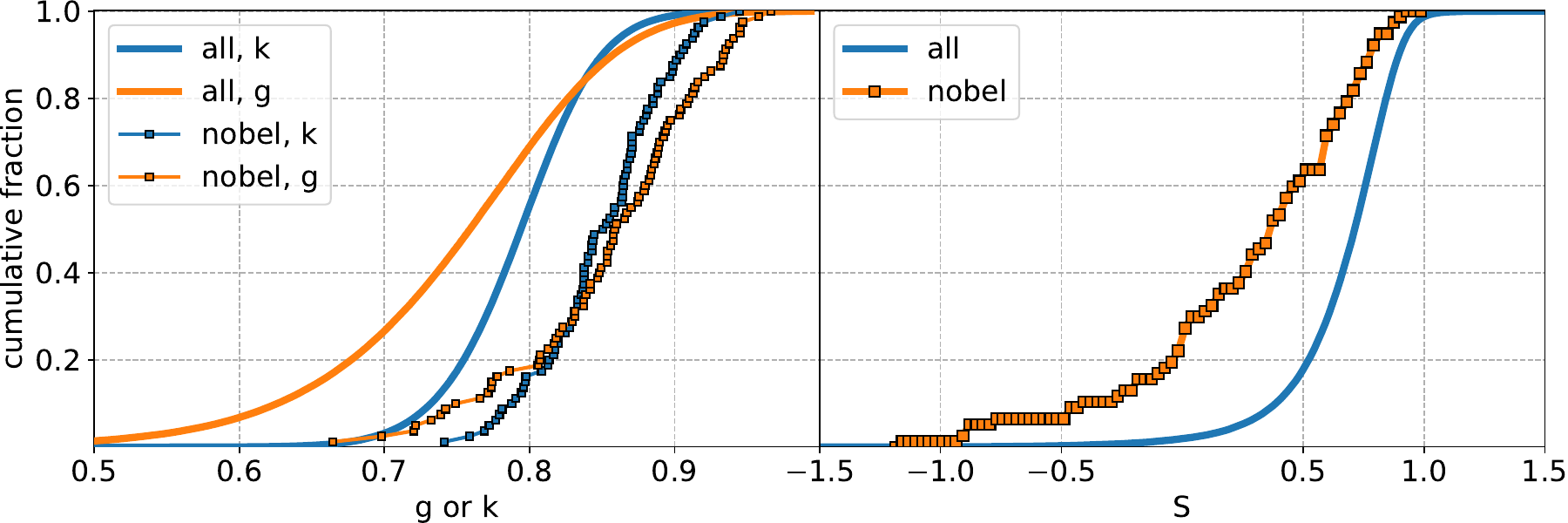}
	\caption{Cumulative fraction of scientists as a function of Gini ($g$, orange) and Kolkata ($k$, blue) indices (\textbf{left}), and Shannon differential entropy $S$ (\textbf{right}). Nobel laureates are highlighted separately in both panels. The results indicate that Nobel laureates on average exhibit higher citation inequality and lower entropy compared to the general population.}
	
	\label{cumul_kg}
\end{figure}

%%%%%%%%%%%%%%%%%%%%%%%%%%%%%%%%%%%%%%%%%%%%%%%%%%%%%%%%%
%\vspace{-6pt}
%%%%%%%%%%%%%%%%%%%%%%%%%%%%%%%%%%%%%%%%%%%%%%%%%%%%%%%%%%

We have also calculated the $Q$-factor from the citation data. While the most probable value of $Q$ for all scientists is around  20, for the Nobel laureates, it is significantly higher. Of course, like $h$, $Q$ also does not have an upper bound.

In addition to the above indices, we computed the Shannon differential entropy based on the normalized citation distributions of individual researchers. The entropy values for Nobel laureates tend to be systematically lower than those of the general population, as shown in the right panel of Figure%MDPI: Wrong Figure ciation number: Figure 4”, “prev number is Figure 6”. Figure 3’s ciation should before Figure 4’s ciation. Please revise and ensure the first citation of each Figure appears in numerical order.
~\ref{cumul_kg} (see also middle panel of Figure~\ref{boxplot}). This indicates that Nobel laureates tend to have citation distributions with lower disorder, as quantified by entropy, which may reflect a more concentrated impact across their publications.

It is worth noting the initially counterintuitive observation that the entropy associated with Nobel laureates tends to assume lower values (often negative) compared to that of the general population, while their inequality measures (e.g., the Gini index) are typically higher. One might expect that higher inequality would correlate with higher entropy. However, if we assume that the underlying distribution follows a Tsallis--Pareto form, this apparent paradox is resolved. For this distribution, both the Gini index and the differential Shannon entropy can be calculated analytically. Specifically, the Gini index is given by \( G = \frac{b}{2b - 1} \), and the differential Shannon entropy by \( S = \frac{1 + b}{b} - \log\left(\frac{b}{b - 1}\right) \). The former is a monotonically decreasing function of \( b \), while the latter increases monotonically, thereby explaining the observed behavior. Notably, the entropy becomes negative for \( b \lesssim 1.19 \), which also aligns with the observed behavior.

We have tabulated the inequality indices for some of the Nobel laureates (2020--2024) in Table \ref{tab1} of Appendix~\ref{AppendixAAA}, where the narrow range of values for $g,k$ for the individual laureates are apparent.

%%%%%%%%%%%%%%%%%%%%%%%%%%%%%%%%%%%%%%%%%%%%%%%%%%%%%%%%%

%%%%%%%%%%%%%%%%%%%%%%%%%%%%%%%%%%%%%%%%%%%%%%%%%%%%%%%%%%

In view of the relative abundance of the population near $g\approx k$ for all scientists, we first establish that, on average, the inequalities are even higher among the Nobel laureates. The left panel of Figure~\ref{cumul_kg} shows the plot of the cumulative fraction of all scientists as functions of $g$ and $k$ and it is then compared with the same plot for the Nobel laureates. It is clear that on average the inequality of the citations is higher among the Nobel laureates. This point in itself is not a conclusive statement on the effectiveness of $g$ and $k$ as indicators on excellence, since the average values of $g, k, Q, h, b$ and $S$ are all significantly different for the Nobel laureates, when compared with the overall data (see Figure~\ref{boxplot}).

\begin{figure}
	
	\includegraphics[width=0.95\linewidth]{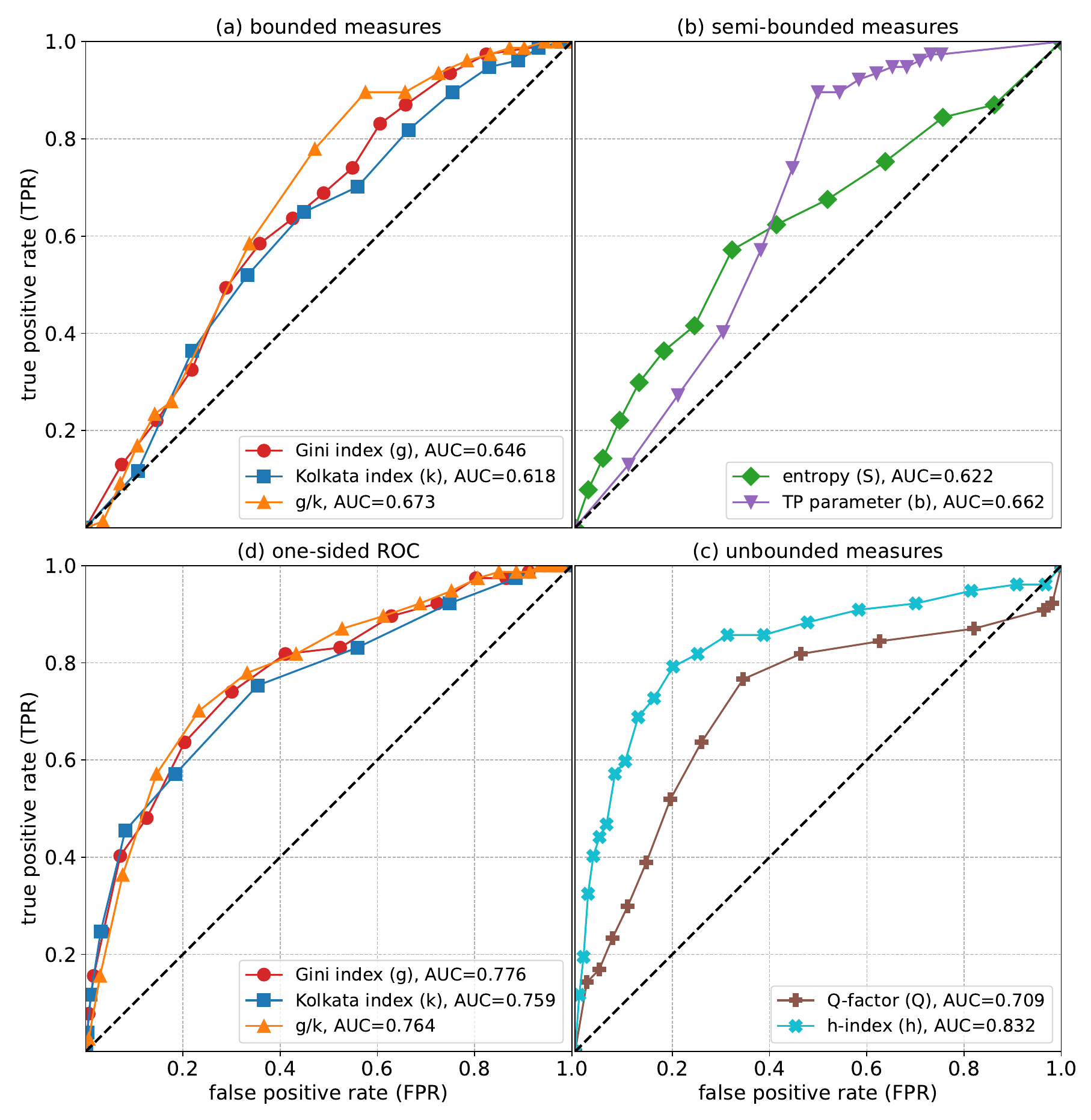}
	\caption{The ROC %MDPI: Please add the explantion of dotted line.
 curves~\cite{roc1,roc2} are shown as follows: {(\textbf{a}) for the bounded measures, viz.,} Gini $g$, Kolkata $k$, and also, for the ratio $g/k$; {(\textbf{b}) for the semi-bounded measures, viz., the entropy $S$ and TP parameter $b$; (\textbf{c}) for the unbounded measures, viz., the Hirsch index $h$ and the $Q$ factor; and (\textbf{d}) the one-sided ROC curves for the bounded measures.} The straight line (45 degree) indicates the complete random process of segregation. The area under the curve (AUC) of the ROC curves and this straight line gives a measure of the overall efficiency of any quantity in distinguishing Nobel laureates from \mbox{the others.}}
	\label{ROC}
\end{figure}

A more standardized measure for the segregation of the Nobel laureates would be to construct a Receiver Operation Characteristic (ROC) curve~\cite{roc1} for each of these indices (or combinations thereof), such that an objective quantification of their relative success in distinguishing the Nobel laureates could be compared. In general, an ROC curve is a plot of True Positive Rate (TPR) with the False Positive Rate (FPR) in an attempted segregation as the parameter in question is gradually varied (see~\cite{roc2} for a review). In the present context, suppose we want to quantify the usefulness of any quantity $J$ in performing the classification. We then measure the average of that quantity for all Nobel laureates. Then, we consider the standard deviation of $J$ among the Nobel laureates and calculate the ratio of the fraction of Nobel laureates within the range $\langle J\rangle-\Delta J$ to $\langle J\rangle+\Delta J$. Subsequently, we calculate the ratio of other scientists within the exact same range, where $\Delta J = n \sigma$, with $\sigma$ being the standard deviation of $J$, and $n$ is a number that we gradually increase.  The fraction of Nobel laureates is then the TPR and the fraction of the others is the FPR, and we plot the points calculated for different values of $n$ (see Figure~\ref{ROC}), giving us the ROC curve. {We also calculate the one-sided ROC curves, for which we  set a threshold of $J$th value for an index, whereby the fraction of Nobel laureates above the threshold gives the TPR and the fraction of the other scientists falling above the threshold gives the FPR.}

The ROC curve, by definition, is bounded between (0, 0) and (1, 1). Clearly, a 45-degree line is when the classifier works no better than random choices. The area under the curve (AUC) between the ROC and 45-degree line is then a measure of the performance of the classifier. This is then a standardized measure for any quantity, particularly the inequality indices, even though the quantities individually may have different ranges of their own. In this way, we can have a quantification of the comparison of performances of different scientometric {metrics} %EE: Please check intended meaning is retained.­
in segregating the Nobel laureates. 

In what follows, {we carry out systematic analysis in terms of the abovementioned ROC curves for all the parameters (inequality indices and the entropy-related parameters): $g$, $k$, $h$, $Q$, $b$ and $S$. We also look at the one-sided ROC curve, but in this case, we only keep the bounded inequality indices{. For unbounded parameters, setting the threshold is dependent upon the present range of the parameter, which in turn may depend upon the time elapsed since {receiving the }award, {particularly} for Nobel laureates}.%EE: Please check intended meaning is retained.­
}

As shown in Figure~\ref{ROC}, we compare {all quantities: \( g \), \( k \), their ratio \( g/k \), entropy $S$, TP parameter $b$ and the one-sided ROC for the bounded measures}. Each of these measures serves as a relatively good indicator of citation inequality. However, the selection of a specific metric depends on the acceptable trade-off between false and true positives. Notably, \( g/k \) appears to perform slightly better than \( g \) and \( k \) individually, suggesting that the combined measure may capture inequality patterns more effectively.
{Also, the one-sided ROC plots perform very well compared to the other ROC plots.}

\subsection{Inequality in Olympic Medals}
Finally, we turn to the question of the inequality of Olympic medals received by various countries in the summer and winter versions of the games over the period of 1896--2024 and 1922--2022, respectively. As mentioned before, in this case as well, a high level of competition is present. In the similar way as described before, one can construct the Lorenz curve and then calculate the inequality indices $g$ and $k$ for the medals received by different countries in a given year. 

\begin{figure}
	
	\includegraphics[width=0.9\linewidth]{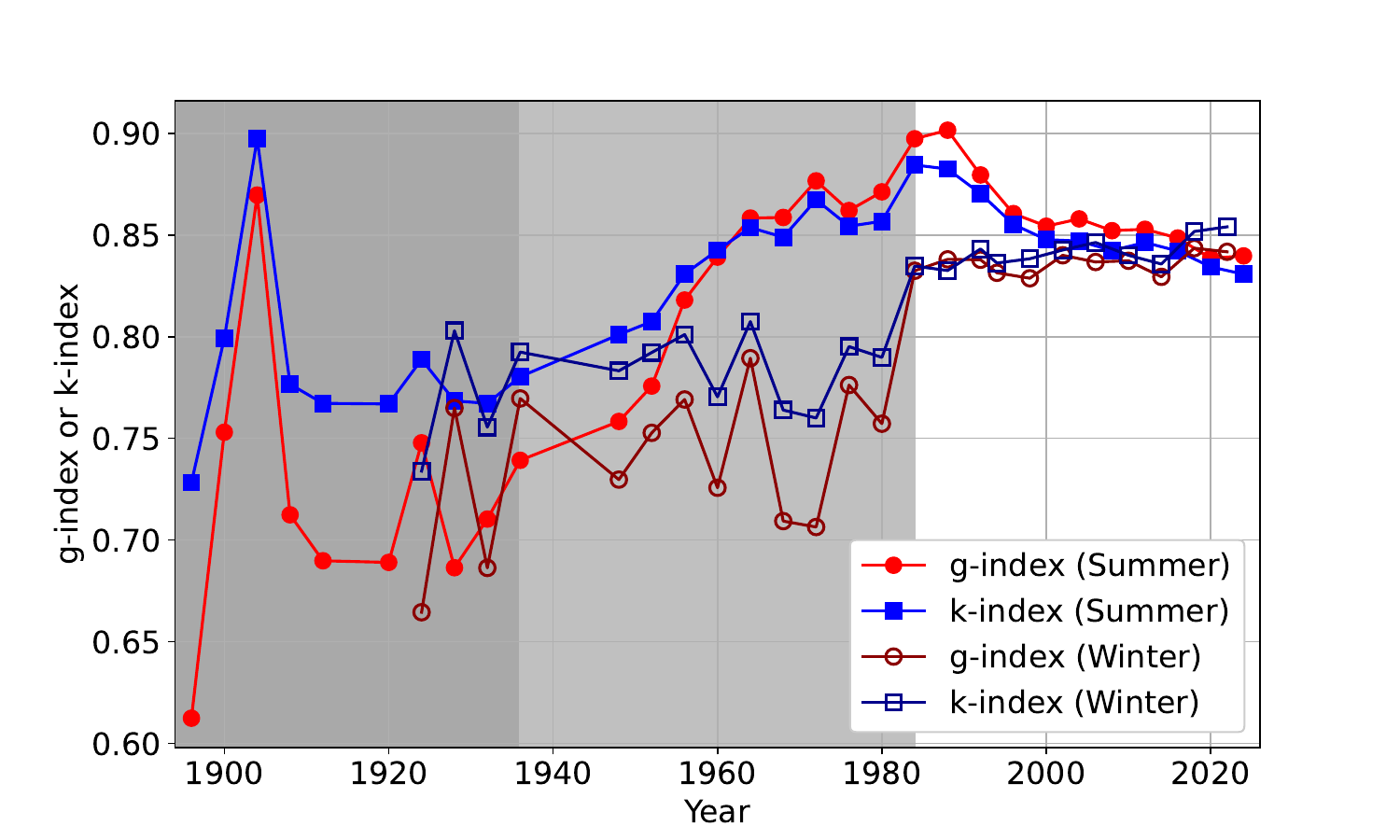}
	\caption{Event-wise (four-yearly) %MDPI: Please add the explantion of Light gray and dark gray.
 variations of the Gini ($g$) and Kolkata ($k$) indices for the number of Olympic medals won by the participating countries are presented for both Summer (filled markers) and Winter (empty markers) Games from 1896 to 2024 (see Tables \ref{tab-Olympic} and \ref{tab-Olympic-win}). Shaded regions indicate years with fewer than 50 participating countries, with lighter gray representing Winter Olympics and darker gray representing Summer Olympics.}
	\label{sum_oly}
\end{figure}

In Figure~\ref{sum_oly}, the time variation of $g$ and $k$ are shown for the Summer Olympic medals. Note that the number of participating countries have changed drastically over the years (see Table \ref{tab-Olympic} in Appendix~\ref{AppendixAAA}). In more recent times, the inequality indices $g$ and $k$ have remained close to each other and around $0.85$. A similar trend can also be observed for the Winter Olympic Games (see Table \ref{tab-Olympic-win} in Appendix~\ref{AppendixAAA}).

These results suggest that irrespective of the underlying mechanism, an emergent inequality of near-universal nature arises for asset accumulation, when there is an unrestricted competition. 

%\vspace{-3pt}

%%%%%%%%%%%%%%%%%%%%%%%%%%%%%%%%%%%%%%%%%%
\section{Discussions and Conclusions}
Reduction in inequality is one of the Sustainable Development Goals that has been established by the United Nations since 2015~\cite{sdg}. However, a little more than Pareto's 80--20 %MDPI: Please check if this should be endash.
 (or self-organized critical) level of social inequality seems inherently tied to social
competence and efficiency. An interesting question would be how far such high-level inequalities can be sustained in different spheres of social dynamics? Such questions are addressable, particularly in the contexts where accumulated asset inequalities are not tied to any dire consequences, or even sometimes a desirable phenomenon. If a steady state can then emerge out of the collective behavior of the competing individuals, it is then useful to inquire what the characteristics of such agents who accumulate a higher amount of assets are. 

In this work, we tried to proceed along that direction, particularly in the context of scholarly citations among the papers of individual researchers and the Olympic medals (summer and winter) received by different countries in a year. The two situations are slightly different from one another.
In the first case, i.e., the inequality of scholarly citations of different papers of many researchers, we note that researchers who are widely recognized as successful (Nobel laureates) have a higher inequality in the  citations of their individual papers. More particularly, the Nobel laureates are disproportionately concentrated near the higher values of the inequality indices (Gini $g$ and Kolkata $k$ indices) when labeled with the inequality indices associated with them.  We have made a quantitative analysis regarding the effectiveness of using the $g,k$ and $g/k$ indices in distinguishing Nobel laureates from the other scientists (see Figure~\ref{ROC}). {We have also studied the effectiveness, using the ROC and variants of it, for several other measures, such as the entropy $S$, TP parameter $b$, $Q$ factor and Hirsch index $h$}. It then suggests that inequality in this context could be a useful indicator of excellence.
{This is in line with the observation that scientific impact likely requires more than one index for quantification~\cite{3d_index,born}}. 

{As mentioned before, the parameters we study can be divided broadly into three groups depending on their domain of values, bounded ($g$, $k$), semi-bounded ($S$, $b$) and unbounded ($Q,h$). }
Post-award, the unbounded indices can be disproportionately affected by the scientist's recognition, whereas bounded and semi-bounded measures may remain more stable. Further quantitative analysis is required to validate this observation by examining the time evolution of these indicators and comparing pre- and post-award behavior.
We also fitted the citation count distributions of individual researchers using the mean-normalized Tsallis--Pareto (Lomax II) distribution. A clear distinction was observed in the fitted shape parameter \( b \), with Nobel laureates consistently exhibiting values closer to the lower limit \( b \to 1 \). This suggests that they lie near the boundary of typical citation patterns, in line with their position at the frontier of scientific excellence. Additionally, the Shannon differential entropy computed from normalized citation distributions was found to be systematically lower for Nobel laureates, indicating a more ordered structure in their citation profiles.

It would also be valuable to explore the entropic properties of the Kolkata index, particularly through the maximum of gintropy, and examine its connection to the differences between Nobel laureates and other scientists.

We then also look at the case of inequality in winning the Olympic medals (both in the Summer and Winter Olympics) among the participating countries. Note that the Olympic Games is already the most competitive athletic event. Therefore, arguing along the same lines as above, one would expect the medal distribution to be highly unequal, which is exactly what we find, especially when the number of participating countries is large. The data on the Olympic medal inequalities should then be compared with the data points for the Nobel laureates in the citation case. {{In contrast,} the data for the other scientists {may resemble} the medal tally of less-competitive sporting events, where inequality is not as pronounced as in the Olympics.} %EE: Please check intended meaning is retained.­
It is difficult to conceive of such an example in a similar scale of participation, and we did not investigate along this line. 

Finally, a couple of more points need to be noted for the citation analysis. Firstly, the number of Nobel laureates is restricted by other conditions than just excellence. There are only a fixed maximum number of laureates in a field in a year, which is not awarded posthumously{, among other factors. }%EE: Please check intended meaning is retained.­
This means that there could be many scientists in the range of citation inequality as that of the Nobel laureates {who are just as excellent.} %EE: Please check intended meaning is retained.­
Therefore, the false positive we mentioned earlier is not necessarily ``false'' in terms of distinguishing excellent researchers. A more detailed analysis of the scientists who fall in this range could be a fruitful future direction of research. Finally, an intuitive understanding as to why such citation inequalities are seen for Nobel laureates could be argued in the following way: If many papers of a researcher receive a similar level of (high) citations, a more probable explanation for that could simply be the higher rate of publications in their particular field rather than all those papers being outstanding. This situation would result in a high $h$-index but low $g,k$ and $Q$. We do not, however, see this for Nobel laureates. In this case, the works awarded Nobel prizes would have received much more attention (even prior to the award) than other works of the laureates. There could of course be individual exceptions.  The question of not observing extreme inequality ($g,k$ near unity) for Nobel laureates is more subtle. It could be expected that except for the Nobel-winning work, an outstanding researcher such as a Nobel laureate would have made some other important contributions as well that would receive good citations, thereby preventing the inequality to reach an extreme level.

In conclusion, we find that emergent inequality is a good indicator of high competitiveness and excellence. We demonstrate this through extensive data analysis for scholarly citations and Olympic medal tallies over the years.

\acknowledgments{The work of M.J. and Z.N. was supported by the project “A better understanding of socio-economic systems using quantitative methods from Physics” funded by the European Union – NextgenerationEU and the Romanian Government, under National Recovery and Resilience Plan for Romania, contract no 760034/23.05.2023, code PNRR-C9-I8-CF255/29.11.2022, through the Romanian Ministry of Research, Innovation and Digitalization, Romania, within Component 9, Investment I8.}

%\conflictsofinterest{The authors declare no conflicts of interests.} 

%%%%%%%%%%%%%%%%%%%%%%%%%%%%%%%%%%%%%%%%%%
%% Optional

%% Only for journal Encyclopedia
%\entrylink{The Link to this entry published on the encyclopedia platform.}

\appendix
\label{AppendixAAA}
\section[\appendixname~\thesection]{}
In this Appendix we list the values of inequality indices of citations of some of the Nobel laureates (between 2020-2024), the inequality of medal tallies in the summer and winter Olympic games during the periods  1896-2024 and 1924-2022 respectively.
\begin{table}[ht]
	\centering
	\scriptsize
	%\begin{adjustbox}{width=1\textwidth}
	\begin{tabular}{|c|c|c|c|c|c|c|c|c|}
		\hline
		Name  & Year & Sub. & NP & NC & h & Q & g & k \\ \hline
		\hline 
		Alain Aspect & 2022 & phys &757 &40141 &75 &122.893&0.9337 &0.8969 \\ \hline 
		Anne L Huillier  & 2023 & phys &504 &34777 &84 &77.1796&0.8469 &0.8377 \\ \hline 
		Anton Zeilinger & 2022 & phys &1098 &113609 &147 &71.3036&0.8897 &0.8707 \\ \hline 
		Ardem Patapoutian & 2021 & med &184 &50057 &87 &11.2758&0.7387 &0.7793 \\ \hline 
		Benjamin List & 2021 & chem &337 &48209 &101 &28.4021&0.7492 &0.7878 \\ \hline 
		Carolyn R. Bertozzi & 2022 & chem &1000 &97852 &150 &37.1544&0.8072 &0.8126 \\ \hline 
		%Claudia Goldin & 2023 & eco &367 &50645 &85 &41.4032&0.8811 &0.8602 \\ \hline 
		Daron Acemoglu & 2024 & eco &1268 &256141 &178 &94.1268&0.9101 &0.8885 \\ \hline 
		David Baker & 2024 & chem &2503 &184477 &220 &62.2481&0.8388 &0.8372 \\ \hline 
		David Card & 2021 & eco &663 &99436 &114 &43.0442&0.8924 &0.8763 \\ \hline 
		David W.C. MacMillan & 2021 & chem &560 &81172 &130 &62.8993&0.8297 &0.8308 \\ \hline 
		Demis Hassabis & 2024 & chem &162 &194682 &96 &28.6888&0.8480 &0.8444 \\ \hline 
		Emmanuelle Charpentier  & 2020 & chem &269 &59389 &62 &94.3811&0.9315 &0.9012 \\ \hline 
		Ferenc Krausz & 2023 & phys &1117 &87620 &129 &86.0893&0.8860 &0.8644 \\ \hline 
		Gary Ruvkun & 2024 & med &343 &71629 &111 &34.5301&0.8130 &0.8201 \\ \hline 
		Geoffrey Hinton & 2024  & phys &724 &905074 &188 &138.225&0.9406 &0.9124 \\ \hline 
		Giorgio Parisi & 2021 & phys &1115 &108250 &134 &117.507&0.8419 &0.8334 \\ \hline 
		Guido W. Imbens & 2021 & eco &348 &110070 &101 &26.0981&0.8747 &0.8674 \\ \hline 
		James Robinson & 2024 & eco &808 &123197 &102 &123.98&0.9452 &0.9138 \\ \hline 
		Jennifer A. Doudna & 2020 & chem &841 &143756 &159 &121.594&0.8586 &0.8431 \\ \hline 
		John F. Clauser & 2022 & phys &133 &21317 &32 &64.0487&0.9452 &0.9164 \\ \hline 
		John Hopfield & 2024 & phys &303 &92855 &94 &93.1922&0.8936 &0.8672 \\ \hline 
		John Jumper & 2024 & chem &72 &58763 &29 &40.8523&0.9250 &0.9038 \\ \hline 
		Joshua D. Angrist & 2021 & eco &400 &101784 &91 &107.176&0.9159 &0.8884 \\ \hline 
		Katalin Kariko & 2023 & med &222 &29678 &66 &20.0473&0.8373 &0.8350 \\ \hline 
		Michael Houghton & 2020 & med &533 &60722 &106 &89.2609&0.8418 &0.8357 \\ \hline 
		Morten Meldal & 2022 & chem &409 &31525 &68 &136.415&0.8208 &0.8135 \\ \hline 
		Moungi G. Bawendi & 2023 & chem &971 &173369 &187 &71.702&0.8315 &0.8308 \\ \hline 
		Paul R. Milgrom & 2020 & eco &383 &116246 &85 &41.2521&0.9085 &0.8905 \\ \hline 
		Robert B. Wilson & 2020 & eco &285 &35042 &58 &41.9753&0.8824 &0.8707 \\ \hline 
		Simon Johnson & 2024 & eco &849 &90468 &65 &178.742&0.9666 &0.9449 \\ \hline 
		Svante Paabo & 2022 & med &581 &144899 &177 &66.567&0.7777 &0.7975 \\ \hline 
		Syukuro Manabe & 2021 & phys &290 &48232 &89 &36.7611&0.8228 &0.8244 \\ \hline 
		Victor Ambros & 2024 & med &180 &71607 &71 &47.1312&0.8842 &0.8661 \\ \hline 
		\hline
	\end{tabular}
	\caption{The table shows various statistical values for Nobel laureates who have Google Scholar profiles and won a Nobel Prize between 2020 and 2024.}
	\label{tab1}
\end{table}

\begin{table}
	\centering
	\scriptsize
	\begin{tabular}{|c|c|c|c|c|c|c|}
		\hline
		Year & Participating & Total Medals & h  &  Q & g & k \\ 
		&  &  &   &   &  &  \\ 
		& countries  &  &   &   &  &  \\ \hline
		\hline
		1896&14&122&6&5.78&0.6124&0.7283\\ \hline
		1900&26&284&6&10.65&0.7530&0.7992\\ \hline
		1904&12&280&4&11.51&0.8696&0.8973\\ \hline
		1908&22&324&8&10.36&0.7124&0.7766\\ \hline
		1912&28&317&8&5.95&0.6898&0.7672\\ \hline
		1920&29&449&11&6.35&0.6890&0.7669\\ \hline
		1924&44&392&10&11.36&0.7478&0.7888\\ \hline
		1928&46&356&10&7.39&0.6864&0.7684\\ \hline
		1932&37&370&10&11.30&0.7103&0.7672\\ \hline
		1936&49&422&12&11.97&0.7392&0.7803\\ \hline
		1948&59&443&12&11.38&0.7583&0.8010\\ \hline
		1952&69&459&11&11.59&0.7757&0.8074\\ \hline
		1956&72&451&12&15.86&0.8180&0.8308\\ \hline
		1960&83&461&10&18.77&0.8391&0.8425\\ \hline
		1964&93&504&12&17.90&0.8583&0.8536\\ \hline
		1968&112&527&13&22.94&0.8586&0.8489\\ \hline
		1972&121&600&13&20.13&0.8766&0.8673\\ \hline
		1976&92&613&12&18.96&0.8620&0.8543\\ \hline
		1980&80&631&12&25.03&0.8712&0.8567\\ \hline
		1984&140&688&13&35.66&0.8973&0.8844\\ \hline
		1988&159&739&14&28.58&0.9015&0.8824\\ \hline
		1992&169&815&16&23.36&0.8795&0.8703\\ \hline
		1996&197&842&16&23.75&0.8605&0.8552\\ \hline
		2000&199&927&16&20.06&0.8543&0.8477\\ \hline
		2004&201&926&16&22.03&0.8579&0.8470\\ \hline
		2008&204&958&16&23.97&0.8521&0.8423\\ \hline
		2012&204&960&16&22.21&0.8528&0.8463\\ \hline
		2016&207&972&17&25.89&0.8485&0.8421\\ \hline
		2020&206&1080&17&21.66&0.8385&0.8343\\ \hline
		2024&207&1044&15&25.10&0.8397&0.8307\\ \hline
		\hline 
	\end{tabular}
	\caption{The table shows various statistical values for Olympic medals (Summer Olympics) won by different countries from 1896 to 2024. }
	\label{tab-Olympic}
\end{table}

\begin{table}
	\centering
	\begin{tabular}{|c|c|c|c|c|c|c|}
		\hline
		Year & Participating & Total Medals & h  &  Q & g & k \\ 
		&  countries  &  &   &   &  &  \\ \hline
		\hline
		1924 & 16 & 49 & 4 & 5.55 & 0.6645 & 0.7339 \\ \hline
		1928 & 25 & 41 & 4 & 9.15 & 0.7649 & 0.8029 \\ \hline
		1932 & 17 & 42 & 3 & 4.86 & 0.6863 & 0.7555 \\ \hline
		1936 & 28 & 51 & 4 & 8.24 & 0.7696 & 0.7924 \\ \hline
		1948 & 28 & 74 & 6 & 5.30 & 0.7297 & 0.7832 \\ \hline
		1952 & 30 & 67 & 5 & 7.16 & 0.7527 & 0.7922 \\ \hline
		1956 & 32 & 72 & 6 & 7.11 & 0.7691 & 0.8010 \\ \hline
		1960 & 30 & 83 & 6 & 7.59 & 0.7257 & 0.7704 \\ \hline
		1964 & 36 & 103 & 7 & 8.74 & 0.7894 & 0.8074 \\ \hline
		1968 & 37 & 106 & 7 & 4.89 & 0.7093 & 0.7639 \\ \hline
		1972 & 35 & 105 & 6 & 5.33 & 0.7064 & 0.7600 \\ \hline
		1976 & 37 & 111 & 6 & 9.00 & 0.7762 & 0.7951 \\ \hline
		1980 & 37 & 115 & 6 & 7.40 & 0.7572 & 0.7898 \\ \hline
		1984 & 49 & 117 & 6 & 10.47 & 0.8324 & 0.8347 \\ \hline
		1988 & 57 & 138 & 7 & 11.98 & 0.8380 & 0.8325 \\ \hline
		1992 & 64 & 171 & 7 & 9.73 & 0.8378 & 0.8431 \\ \hline
		1994 & 67 & 183 & 8 & 9.52 & 0.8314 & 0.8362 \\ \hline
		1998 & 72 & 205 & 10 & 10.19 & 0.8287 & 0.8383 \\ \hline
		2002 & 77 & 237 & 9 & 11.70 & 0.8400 & 0.8426 \\ \hline
		2006 & 80 & 252 & 11 & 9.21 & 0.8367 & 0.8463 \\ \hline
		2010 & 82 & 258 & 10 & 11.76 & 0.8373 & 0.8400 \\ \hline
		2014 & 88 & 284 & 10 & 8.68 & 0.8294 & 0.8357 \\ \hline
		2018 & 92 & 307 & 12 & 11.69 & 0.8435 & 0.8517 \\ \hline
		2022 & 91 & 327 & 13 & 10.30 & 0.8417 & 0.8540 \\ \hline
		\hline 
	\end{tabular}
	\caption{The table shows various statistical values for Olympic medals (Winter Olympics) won by different countries during the period 1924-2022.}
	\label{tab-Olympic-win}
\end{table}


\begin{thebibliography}{999}
% Reference 1 (Journal Article)
\bibitem[Stauffer(2016)]{stauffer}
Stauffer, D. Income Inequality in the 21st Century—A Biased Summary of Piketty’s Capital in the Twenty-First Century. {\em Int. J. Mod. Phys. C} {\bf 2016}, {\em 27}, 1630001.

\bibitem[Pareto(1896)]{pareto}
Pareto, V. Cours d’\'{e}conomie politique. {\em Political Sci. Q.} {\bf 1896}, {\em 11}, 750–751.

\bibitem[Dubinsky and Hansen(1982)]{manage}
Dubinsky, A.J.; Hansen, R.W. Improving Marketing Productivity: The 80/20 Principle Revisited. {\em Calif. Manag. Rev.} {\bf 1982}, {\em 25}, 96--105.

\bibitem[Woolhouse et al.(1997)]{gu1}
Woolhouse, M.; Dye, C.; Etard, J.; Smith, T.; Charlwood, J.; Garnett, G.; Hagan, P.; Hii, J.; Ndhlovu, P.; Quinnell, R.; et al. Heterogeneities in the Transmission of Infectious Agents: Implications for the Design of Control Programs. {\em Proc. Natl. Acad. Sci. USA} {\bf 1997}, {\em 94}, 338.

\bibitem[Abeles and Conway(2020)]{gu2}
Abeles, J.; Conway, D.J. The Gini Coefficient as a Useful Measure of Malaria Inequality Among Populations. {\em Malar. J.} {\bf 2020}, \mbox{{\em 19}, 444.}

\bibitem[Manna et al.(2022)]{manna}
Manna, S.S.; Biswas, S.; Chakrabarti, B.K. Near Universal Values of Social Inequality Indices in Self-Organized Critical Models. {\em Phys. A} {\bf 2022}, {\em 596}, 127121.


\bibitem[Banerjee et al.(2023)]{ijmpc}
Banerjee, S.; Biswas, S.; Chakrabarti, B.K.; Challagundla, S.K.; Ghosh, A.; Guntaka, S.R.; Koganti, H.; Kondapalli, A.R.; \mbox{Maiti, R.;} Mitra, M.; et al. Evolutionary Dynamics of Social Inequality and Coincidence of Gini and Kolkata Indices under Unrestricted Competition. {\em Int. J. Mod. Phys. C} {\bf 2023}, {\em 34}, 2350048.

\bibitem[Bir\'{o} et al.(2023)]{zoltan}
Bir\'{o}, T.S.; Andras, T.; J\'{o}zsa, M.; N\'{e}da, Z. Gintropic Scaling of Scientometric Indexes. {\em Phys. A Stat. Mech.   Its Appl.} {\bf 2023}, \mbox{{\em 618}, 128717.}

\bibitem{nielsen}
  {  Nielsen, M.W.; Andersen,  J.P. 
Global citation inequality is on the rise.\emph{ Proc. Natl. Acad. Sci. USA} \textbf{2021}, \emph{118}, e2012208118}.

\bibitem{dong}
Dong, K.;Wu,  J.; Wang, K. 
{On the inequality of citation counts of all publications of individual authors}.
\emph{J. Inf.} \textbf{2021}, \emph{15}, 101203.



\bibitem{teich}
Teich, E.G.; Kim, J.Z.; Lynn, C.W.; Simon, S.C.; Klishin, A.A.; Szymula, K.P.; Srivastava, P.; Bassett, L.C.;  Zurn, P.; Bassett, D.S.; et al. {Citation inequity and gendered citation practices in contemporary physics}. Nat. Phys. \textbf{2022}, \emph{18}, 1161–1170.

\bibitem{nett}
Nettasinghe, B.; Alipourfard, N.; Krishnamurthy, V.; Lerman, K. {Emergence
of structural inequalities in scientific citation networks}.  \emph{arxiv} \textbf{2021}, arXiv:2103.10944.

\bibitem{crespo}
{Crespo, J.A.; Li, Y.; Ruiz-Castillo, J.  {The measurement of the effect on citation inequality of differences in citation practices across scientific fields}. \emph{PLoS ONE} \textbf{2013},  \emph{8}, e58727.}

\bibitem{kiess}
{Kiesslich, T.; Beyreis, M.; Zimmermann, G.; Traweger, A. {Citation inequality and the Journal Impact Factor: Median, mean, (does it) matter?}~\emph{Scientometrics} \textbf{2021}, \emph{126}, 1249–1269}.

\bibitem[Banerjee et al.(2023)]{entropy}
Banerjee, S.; Biswas, S.; Chakrabarti, B.K.; Ghosh, A.; Mitra, M. Sandpile Universality in Social Inequality: Gini and Kolkata Measures. {\em Entropy} {\bf 2023}, {\em 25}, 735.

\bibitem[Ghosh et al.(2024)]{Qfactor}
Ghosh, A.; Manna, S.S.; Chakrabarti, B.K. Q Factor: A Measure of Competition Between the Topper and the Average in Percolation and in Self-Organized Criticality. {\em Phys. Rev. E} {\bf 2024}, {\em 110}, 014131.



\bibitem[Lorenz(1905)]{lorenz}
Lorenz, M.O. Methods of Measuring the Concentration of Wealth. {\em Publ. Am. Stat. Assoc.} {\bf 1905}, {\em 9}, 209.

\bibitem[Gini(1921)]{gini}
Gini, C. Measurement of Inequality of Incomes. {\em Econ. J.} {\bf 1921}, {\em 31}, 124.



%\bibitem[Bir\'{o} and N\'{e}da(2020)]{gu3}
%Bir\'{o}, T.S.; N\'{e}da, Z. Gintropy: Gini Index Based Generalization of Entropy. Entropy {\bf \hl{2020}}, {\em \hl{22}}, \hl{879}. \url{https://doi.org/10.3390/e22080879}.
%MDPI: Refs. 19 and 25 are duplicated. Please remove duplicated references and rearrange all the references to appear in numerical order. Please ensure that there are no duplicated references.

\bibitem[Biro and Neda(2020)]{Bir2020}
Bir\'{o}, T.S.; N\'{e}da, Z. Gintropy: Gini Index Based Generalization of Entropy. {\em Entropy} {\bf 2020}, {\em 22}, 879.


\bibitem[Ghosh et al.(2014)]{kolkata}
Ghosh, A.; Chattopadhyay, N.; Chakrabarti, B.K. Inequality in Societies, Academic Institutions and Science Journals: Gini and k-Indices. {\em Phys. A} {\bf 2014}, {\em 410}, 30.

\bibitem{muthu}
{Banerjee, S.; Chakrabarti, B.K.; Mitra, M.; Mutuswami, S. Inequality Measures: The Kolkata Index in Comparison With Other Measures. \emph{Front. Phys.} \textbf{2020}, \emph{8}, 562182}.

\bibitem[Hardy(2010)]{pindex}
Hardy, M. Pareto's Law. {\em Math. Intell.} {\bf 2010}, {\em 32}, 38.

\bibitem[Cui et al.(2023)]{kpa1}
Cui, L.; Lin, C.; Huang, X. Kinetic Modeling of Wealth Distribution with Saving Propensity, Earnings Growth, and Matthew Effect. {\em Europhys. Lett.} {\bf 2023}, {\em 143}, 12002.

\bibitem[Lin and Cui(2025)]{kpa2}
Lin, C.; Cui, L. Kinetic Modelling of Economic Markets with Individual and Collective Transactions. {\em arXiv} {\bf 2025}, arXiv:2502.13735.






\bibitem[Hirsch(2005)]{hindex}
Hirsch, J.E. An Index to Quantify an Individual’s Scientific Research Output. {\em Proc. Natl. Acad. Sci. USA} {\bf 2005}, {\em 102}, 16569.

\bibitem{h_rev1}
{Alonso, S.; Cabrerizo, F.J.; Herrera-Viedma, E.; Herrera, F.
{h-Index: A review focused in its variants, computation and standardization for different scientific fields}.
\emph{J. Inf.} \textbf{2009}, \emph{3}, 273--289.}

\bibitem{h_rev2}
{Bihari, A.; Tripathi, S.; Deepak, A.  {A review on h-index and its alternative indices}. \emph{J. Inf. Sci.} {\textbf{2021}, \emph{49}}, 624--665.}


\bibitem[Figshare(2025)]{data1}
Author Citation Data. Available online: \url{https://figshare.com/articles/dataset/Data_for_article_i_Does_Excellence_Correspond_to_Universal_Inequality_Level_i_/28827314} (accessed on 19 April 2025).

\bibitem[Topend Sports(2023)]{data2}
Winter Olympics Medal Tally. Available online: %MDPI: In One Reference cannot have two website link, please change it into two refs, and reorder the reference and it's citation.
 \url{https://www.topendsports.com/events/winter/medal-tally/medal-tables.htm} (accessed on 10 March 2025). Summer Olympics Medal Tally. Available online: \url{https://www.olympics.com/en/olympic-games}.


\bibitem[Ghosh and Chakrabarti(2024)]{soc-citation}
Ghosh, A.; Chakrabarti, B.K. Do Successful Researchers Reach the Self-Organized Critical Point? {\em Physics} {\bf 2024}, {\em 6}, 46–59. \url{https://doi.org/10.3390/physics6010004}.

\bibitem[Yong(2014)]{yong}
Yong, A. A Critique of Hirsch’s Citation Index: A Combinatorial Fermi Problem. {\em Not. Am. Math. Soc.} {\bf 2014}, {\em 61}, 1040–1050.

\bibitem[Tanner and Swets(1954)]{roc1}
Tanner, W.P., Jr.; Swets, J.A. A Decision-Making Theory of Visual Detection. {\em Psychol. Rev.} {\bf 1954}, {\em 61}, 401–409.

\bibitem[Fawcett(2006)]{roc2}
Fawcett, T. An Introduction to ROC Analysis. {\em Pattern Recognit. Lett.} {\bf 2006}, {\em 27}, 861.



\bibitem[United Nations SDGs(2023)]{sdg}
Sustainable Development Goals. Available online: \url{https://sdgs.un.org/goals} (accessed on 10 March 2025).

\bibitem{3d_index}
{Siudem, G.; Żogała-Siudem, B.; Cena, A.; Gagolewski, M. {Three dimensions of scientific impact}. \emph{Proc. Natl. Acad. Sci. USA} {\textbf{2020}, \emph{117}}, 13896--13900, \url{https://doi.org/10.1073/pnas.2001064117}.}

\bibitem{born}
{Bornmann, L.; Mutz, R.; Daniel, H.D. {Are there better indices for evaluation purposes than the h index? A comparison of nine different variants of the h index using data from biomedicine}. \emph{J. Am. Soc. Inf. Sci. Technol.} \textbf{2008}, \emph{59}, 830--837.}
\end{thebibliography}
\end{document}